\documentclass[12pt]{article}
\usepackage{cite}
\usepackage{graphicx}

\let\*=\,
\newcommand{\C}[5]{\overline{\mathcal{C}}_{#1,#2}^{(#3#4),#5}}
\newcommand{\CL}[4]{\overline{\mathcal{C}}_{#1}^{(#2#3),#4}}
\newcommand{\M}[3]{{\mathrm{T}}^{}_{#2#3}}
\newcommand{\nl}{n_l}
\newcommand{\nh}{n_h}
\newcommand{\T}[2]{T_{#1,#2}}
\newcommand{\z}[1]{\zeta_{#1}}
\newcommand{\Pa}[1]{a_{#1}}
\newcommand{\Log}[2]{\log^{#2}({#1})}
\newcommand{\lmsdmus}{l_{m}}
\newcommand{\s}{\hspace{1.1ex}}

\begin{document}

\begin{titlepage}
\noindent
\hfill BNL-HET-08/12\\
\mbox{}
\hfill May 21, 2008\\
\mbox{}

\vspace{0.5cm}
\begin{center}
  \begin{Large}
    \begin{bf}
      Moments of Heavy Quark Current Correlators\\
      at Four-Loop Order in Perturbative QCD
    \end{bf}
  \end{Large}
  \vspace{0.8cm}

  \begin{large}
    Christian Sturm\footnote{sturm@bnl.gov}
  \end{large}
  \vskip .7cm
	 {\small {\em 
	     Physics Department, 
             Brookhaven National Laboratory, \\
             Upton,
	     New York 11973, USA}}
	
	\vspace{4cm}
{\bf Abstract}
\end{center}
\begin{quotation}
  \noindent
  We present the result for the first moment of the scalar and
  axial-vector current correlator in third order of the strong coupling
  constant $\alpha_s$ and give the details of a recent evaluation of the
  pseudo-scalar correlator. The results can be used to reduce the
  theoretical uncertainty due to higher order corrections for the
  determination of fundamental parameters of QCD in the context of
  lattice calculations.
\end{quotation}
\end{titlepage}
%
%
\section{Introduction\label{sec:Introduction}}
The computation of heavy quark current correlators in perturbation
theory is of central interest for many phenomenological applications.
For instance their low energy expansion can be related to moments, which
have been determined with high precision through the calculation of
higher order corrections. Moments of the vector current correlator play
an important role in combination with sum rules and the experimentally
measured ratio
$R(s)=\sigma(e^+e^-\to\mbox{hadrons})/\sigma(e^{+}e^{-}\to\mu^{+}\mu^{-})$. They
can be used to perform a precise charm- and bottom-quark mass
determination. This method has first been suggested in
Ref.~\cite{Novikov:1977dq,Reinders:1984sr} and has been continuously
improved over the years by considering new data from experiment and by
including higher order
corrections~\cite{Kuhn:2001dm,Kuhn:2002zr,Kuhn:2007vp}. However, also
the scalar, pseudo-scalar and axial-vector current correlators can be
employed in this context. For example, moments of the pseudo-scalar
correlator can be used in combination with lattice calculations to
determine fundamental constants of QCD, the strong coupling constant
$\alpha_s$ and the charm-quark mass with high
accuracy~\cite{Lepage:2008}. Furthermore, the first eight three-loop
moments have been
used~\cite{Chetyrkin:1995ii,Chetyrkin:1996cf,Chetyrkin:1997mb,Chetyrkin:1998ix}
as one of several ingredients in order to reconstruct the complete
momentum and mass dependence of the scalar, pseudo-scalar, vector and
axial-vector polarization functions by means of Pad{\'e}-approximations.
The complete mass and momentum dependence is important in the
computation of the $Z$-boson decays, in which a combination of the
vector and axial-vector density enters or in the calculation of
Higgs-boson decays which are related to the scalar or pseudo-scalar
correlator.

Recently the low energy expansion at three-loop order has been extended
and the first $30$ coefficients have been determined for the vector
current and for the remaining scalar, pseudo-scalar and axial-vector
currents in Ref.~\cite{Boughezal:2006uu,Maier:2007yn}, where also
singlet contributions have been taken into account.  In four-loop order
only the first moment of the vector current correlator is fully known at
present\cite{Chetyrkin:2006xg,
Boughezal:2006px}. For double fermionic contributions the first five
moments have been computed~\cite{Czakon:2007qi}. Contributions of the
order $\alpha_s^j\*\nl^{j-1}$ have been calculated to all orders $j$ in
Ref.~\cite{Grozin:2004ez}, where $\nl$ denotes the number of light
quarks, considered as massless. Numerical results for two moments of the
pseudo-scalar correlator were presented in~\cite{Lepage:2008}.

Whereas the lowest moments of the vector correlator have been studied up
to four-loop order already some time ago, the ones of the scalar,
pseudo-scalar and axial-vector current correlators were available up to
three-loop order only.  The purpose of this paper is to provide the
still unknown four-loop QCD corrections to the lowest moments of the
scalar and axial-vector current correlators and to present the details
of the evaluation for the pseudo-scalar correlator, where first
numerical results were presented in~\cite{Lepage:2008}. Our discussion
will be limited to the non-singlet contributions. These results can be
useful in combination with lattice calculations to reduce the error due
to unknown higher order corrections in perturbation theory in the
context of the determination of the strong coupling constant and quark
mass as demonstrated in~\cite{Lepage:2008}. They can also be seen as a
first step towards the evaluation of higher moments. Furthermore
analytical results of the pseudo-scalar correlator are presented.

The techniques used in this work have already been successfully applied
in several other calculations, among which are the calculation of the
matching relation of the strong coupling constant at a heavy quark
threshold up to four-loop order in perturbative
QCD~\cite{Chetyrkin:2005ia,Schroder:2005hy} and the computation of the
four-loop QCD corrections to the $\rho$-parameter arising from top- and
bottom-quark loops
\cite{Schroder:2005db,Chetyrkin:2006bj,Boughezal:2006xk,Faisst:2006sr}.

The outline of this paper is as follows: In
Section~\ref{sec:GeneralNotations} we introduce our notations and
conventions. In Section~\ref{sec:Calculations} we discuss the methods of
calculation and give the results for the first moment of the scalar and
axial-vector current correlators at four-loop order. For completeness we
recall the results for the pseudo-scalar and vector case. Finally in
Section~\ref{sec:DiscussConclude} we close with a brief summary and our
conclusions. The results for the vector and pseudo-scalar correlator are
listed in Appendix~\ref{app:PseudoVector}, those for the moments with
$n=-1$ and $n=0$ in Appendix~\ref{app:Mom0m1}.
%
\section{Generalities and Notation\label{sec:GeneralNotations}}
The polarization functions for the scalar($s$), pseudo-scalar($p$),
axial-vector($a$) and vector($v$) current correlator are defined by 
\begin{eqnarray}
\label{eq:scalarpseudo}
q^2\*\Pi^{\delta}(q^2)\!&\!\!=\!\!&\!
i\!\int\!\!dx e^{iqx}\langle0|Tj^{\delta}(x)j^{\delta}(0)|0\rangle,
\quad\mbox{for}\; \delta\!=\!s,p\\
(q_{\mu}q_{\nu}-q^2g_{\mu\nu})\*
\Pi^{\delta}(q^2)+q_{\mu}q_{\nu}\Pi^{\delta}_{L}(q^2)\!&\!\!=\!\!&\!
i\!\int\!\!dx e^{iqx}\langle0|Tj_{\mu}^{\delta}(x)j_{\nu}^{\delta}(0)|0\rangle,
\quad\mbox{for}\; \delta\!=\!a,v
\label{eq:vectoraxial}
\end{eqnarray} 
with the currents
\[
j^{s}=\overline{\Psi}\Psi,\qquad
j^{p}=i\overline{\Psi}\gamma_5\Psi,\qquad
j^{a}_{\mu}=\overline{\Psi}\gamma_{\mu}\gamma_5\Psi,\qquad
j^{v}_{\mu}=\overline{\Psi}\gamma_{\mu}\Psi.
\]
The low-energy expansion of the polarization functions in
$z=q^2/(2\*m)^2$ is conveniently written as%
\begin{equation}
\label{eq:expand}
\overline{\Pi}^{\delta}(q^2)={3\over16\*\pi^2}\sum_{n=-1}^{\infty}\overline{C}^{\delta}_n\*z^n,
\end{equation}
where the expansion coefficients $\overline{C}^{\delta}_n$ are computed
up to four-loop order in perturbative QCD.  The expansion in the
coupling constant $\alpha_s/\pi$ up to four-loop order is given by
\begin{equation}
\overline{C}^{\delta}_n=\overline{C}^{(0),\delta}_n
                       +\left({\alpha_s\over\pi}\right)\*\overline{C}^{(1),\delta}_n
                       +\left({\alpha_s\over\pi}\right)^2\*\overline{C}^{(2),\delta}_n
                       +\left({\alpha_s\over\pi}\right)^3\*\overline{C}^{(3),\delta}_n
                       +\dots\,.
\end{equation}
We decompose the coefficient $\overline{C}_{n}^{(i),\delta}$
$(i=0,1,2,3)$ for $n>0$ into the non-logarithmic and logarithmic parts
\begin{equation}
  \overline{C}^{(i),\delta}_{n}=\sum_{j=0}^{i}\overline{C}^{(ij),\delta}_{n}\lmsdmus^{j},
\end{equation}
with $\lmsdmus=\log\left({m^2\over\mu^2}\right)$.  Furthermore we
classify the diagrams with respect to the number of closed fermion loops
inserted into a diagram. The symbol $\nl$ denotes the number of light
quarks and the symbol $\nh=1$ denotes a heavy quark with mass
$m$. This decomposition at four-loop order is given by
\begin{equation}
\overline{C}^{(3j),\delta}_{n}\!=
 \C{n}{0 }{3}{j}{\delta}\!\!
+\C{n}{h }{3}{j}{\delta}\nh
+\C{n}{l }{3}{j}{\delta}\nl
+\C{n}{hh}{3}{j}{\delta}\nh^2
+\C{n}{hl}{3}{j}{\delta}\nh\nl
+\C{n}{ll}{3}{j}{\delta}\nl^2.
\end{equation}
The bar indicates that renormalization of $m$, $\alpha_s$ and the
current has been performed in the $\overline{\mbox{MS}}$-scheme. We have
checked that for $n=1$ both scalar and axial-vector polarization
functions $\overline{\Pi}^{\delta}(q^2)$ obey the standard
renormalization group equation(RGE). For the vector current correlator
the longitudinal part $\Pi^{v}_{L}(q^2)$ of the polarization function is
zero due to the vector Ward-identity. The longitudinal part of the
axial-vector correlator obeys the axial
Ward-identity\cite{Broadhurst:1974ng,Broadhurst:1981jk,Chetyrkin:1992xk,Chetyrkin:1996hm,Chetyrkin:1996ia}
\begin{equation}
\label{eq:axialWI}
q^4\*\Pi^{a}_L(q^2)=(2\*m)^2\*q^2\*\Pi^{p}(q^2)+\mbox{contact term.}
\end{equation}
Inserting the expansion of Eq.(\ref{eq:expand}) leads to
\begin{equation}
\label{eq:axialWIExpansion}
\sum_{n=-1}^{\infty}\overline{C}^{a}_{L,n}\*z^n=
\sum_{n=-1}^{\infty}\overline{C}^{p}_n\*z^{n-1}+
\frac{1}{q^4}\*\mbox{contact term.}
\end{equation}
Performing a shift in the summation index $n=k+1$ on the r.h.s. of
Eq.(\ref{eq:axialWIExpansion}) and comparing the coefficients of the
different orders in $z^k$ in both sides leads to
\begin{equation}
\overline{C}^{a}_{L,k}=\overline{C}^{p}_{k+1},
\end{equation}
for $k\ge-1$, which allows to obtain the second moment of the
pseudo-scalar correlator from the calculation of the first moment of the
longitudinal part of the axial-vector correlator. The contact term in
Eq.(\ref{eq:axialWIExpansion}) only contributes to the order $1/z^{2}$.
%
%
%
%
\section{Calculations and Results\label{sec:Calculations}}
\begin{figure}[!ht]
\begin{center}
\begin{minipage}[b]{2cm}
  \begin{center}
    \includegraphics[height=1.5cm,bb=126 332 460 665]{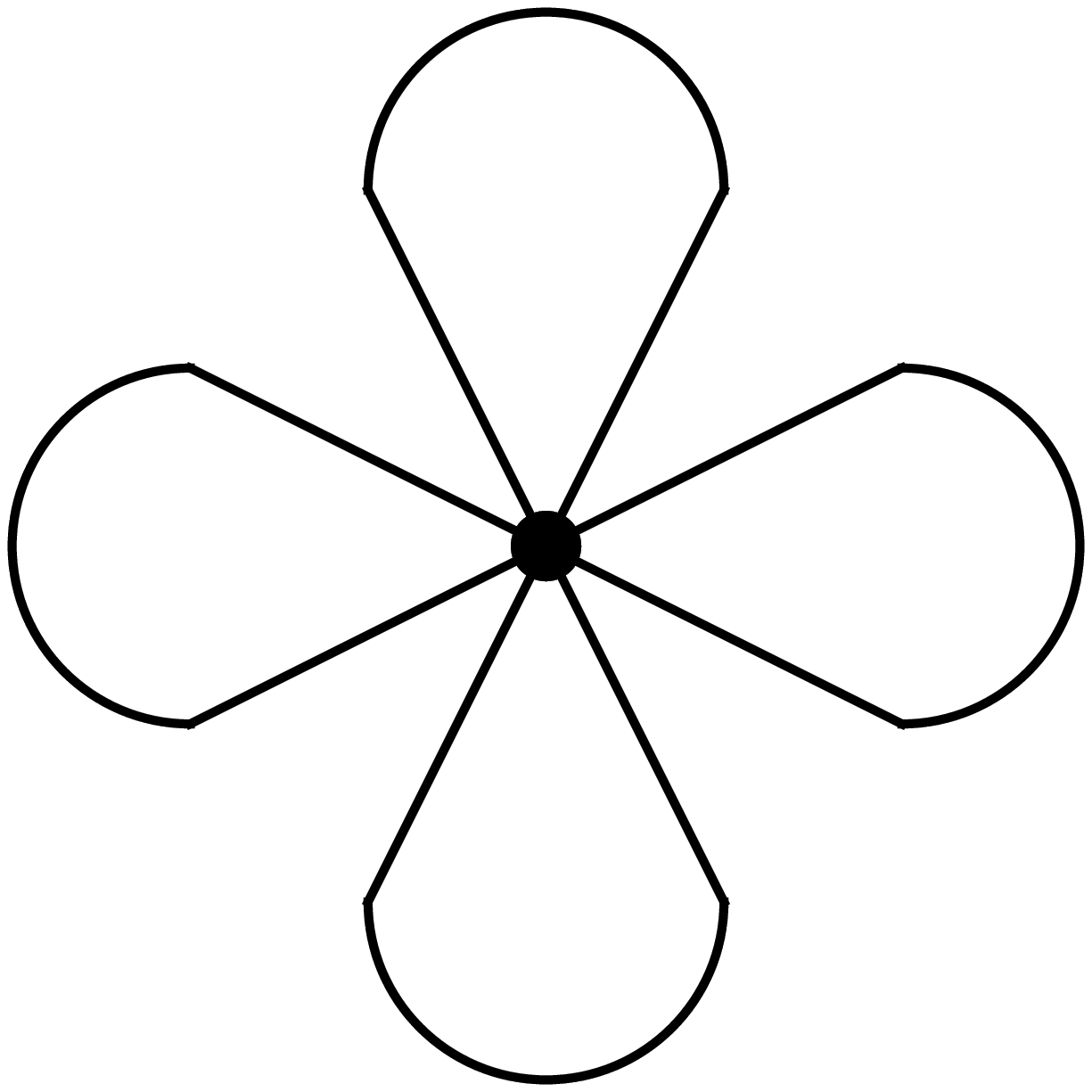}
    $\M{}{4}{1}$
  \end{center}
\end{minipage}
%
%
\begin{minipage}[b]{2cm}
  \begin{center}
    \includegraphics[height=1.5cm,bb=170 320 415 666]{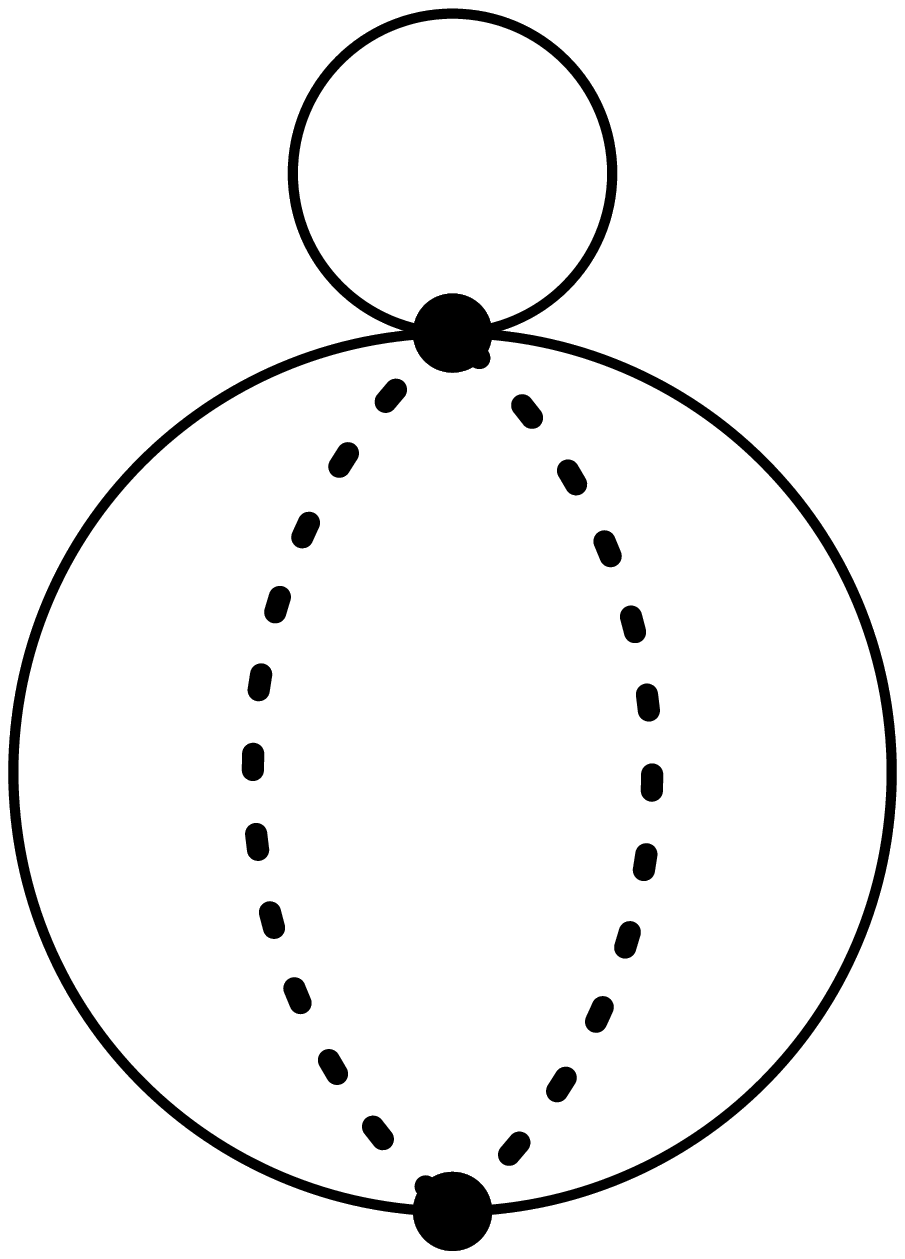}
\\
    $\M{}{5}{1}$
  \end{center}
\end{minipage}
%
%
\begin{minipage}[b]{2cm}
  \begin{center}
    \includegraphics[height=1.5cm,bb=126 320 460 678]{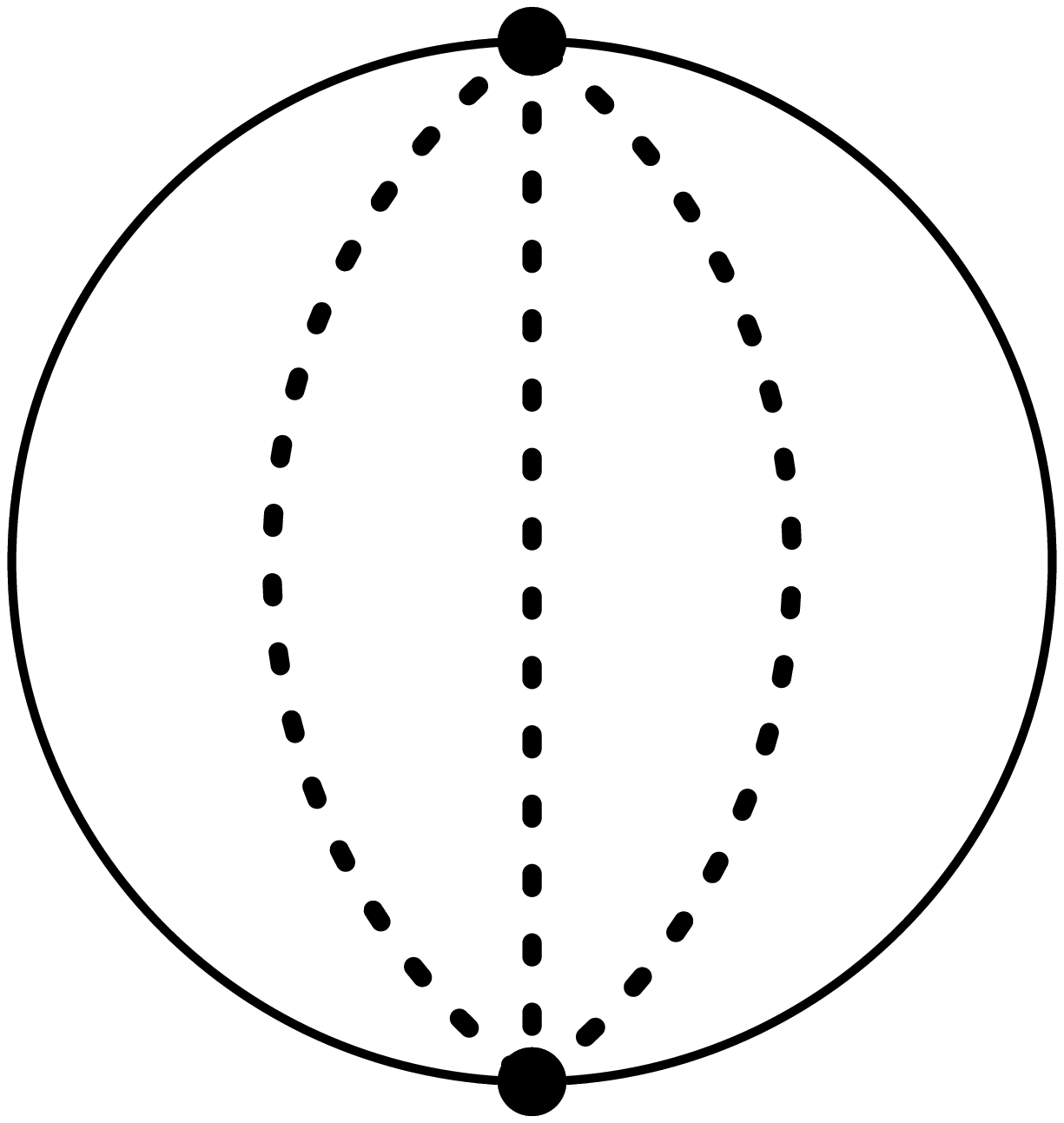}
    $\M{}{5}{3}$
  \end{center}
\end{minipage}
%
%
\begin{minipage}[b]{2cm}
  \begin{center}
    \includegraphics[height=1.5cm,bb=126 320 460 678]{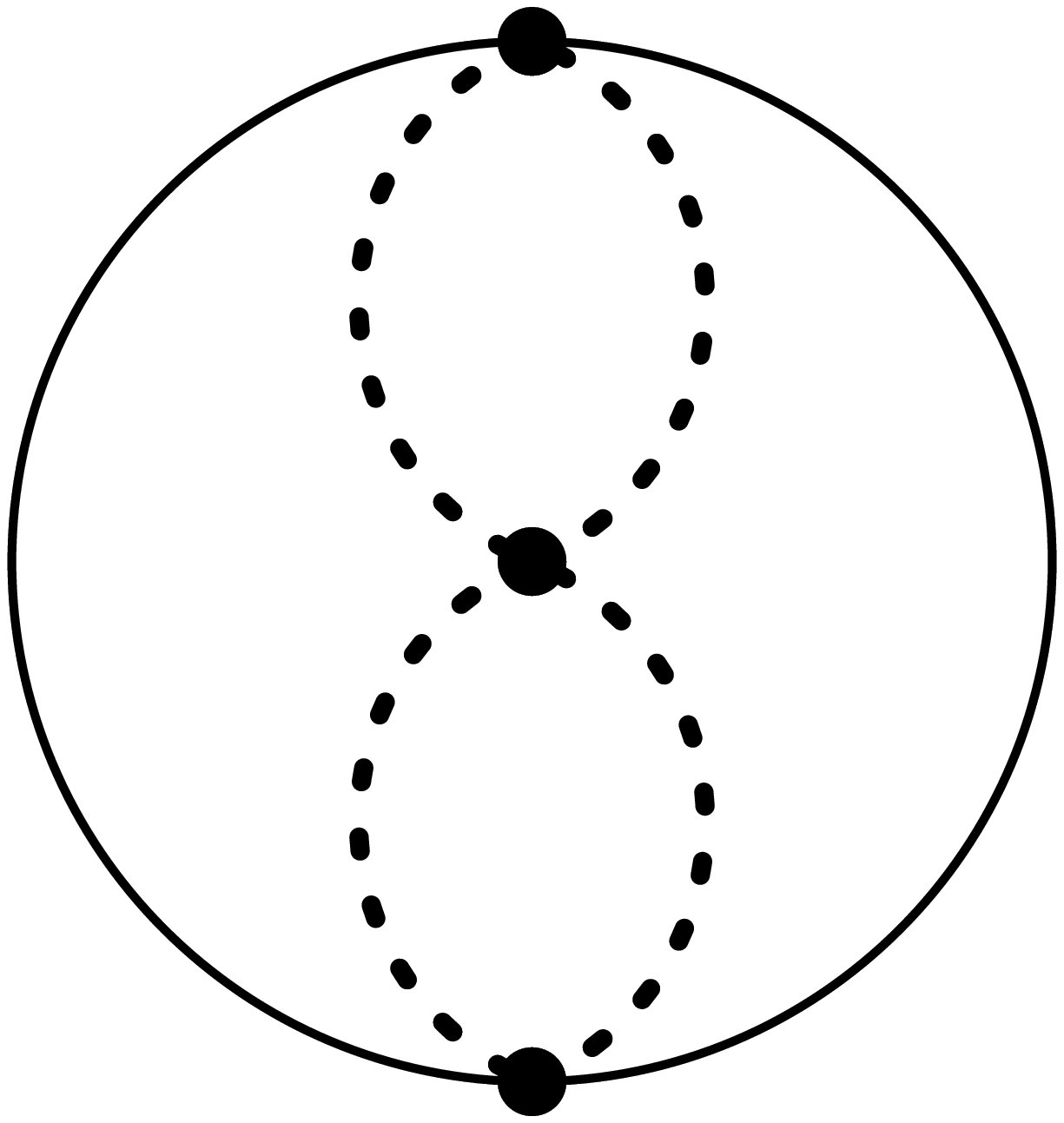}
    $\M{}{6}{3}$
  \end{center}
\end{minipage}
%
%
\begin{minipage}[b]{2cm}
  \begin{center}
    \includegraphics[height=1.5cm,bb=170 320 415 666]{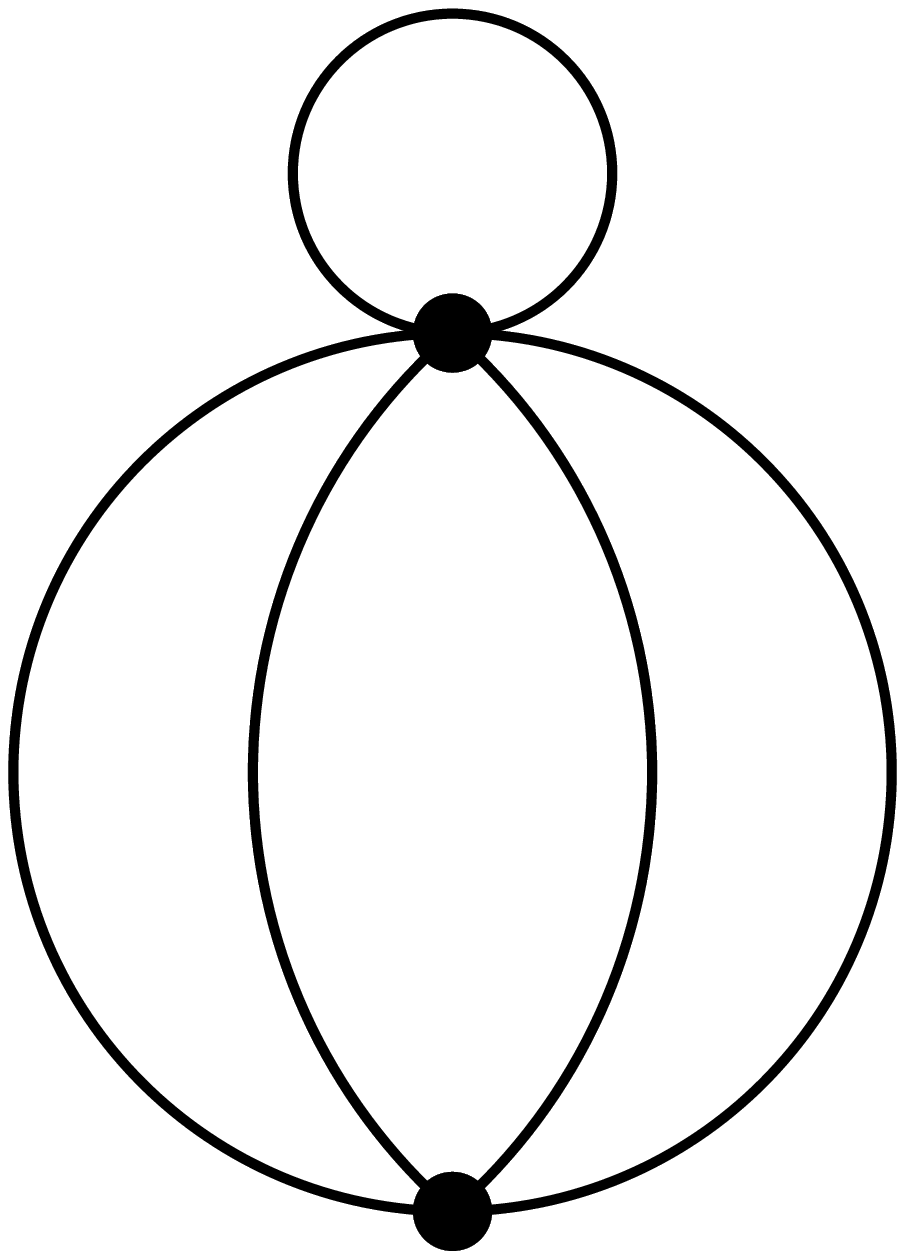}
\\
    $\M{}{5}{2}$
  \end{center}
\end{minipage}
\\
%
%
\begin{minipage}[b]{2cm}
  \begin{center}
    \includegraphics[height=1.5cm,bb=126 320 460 678]{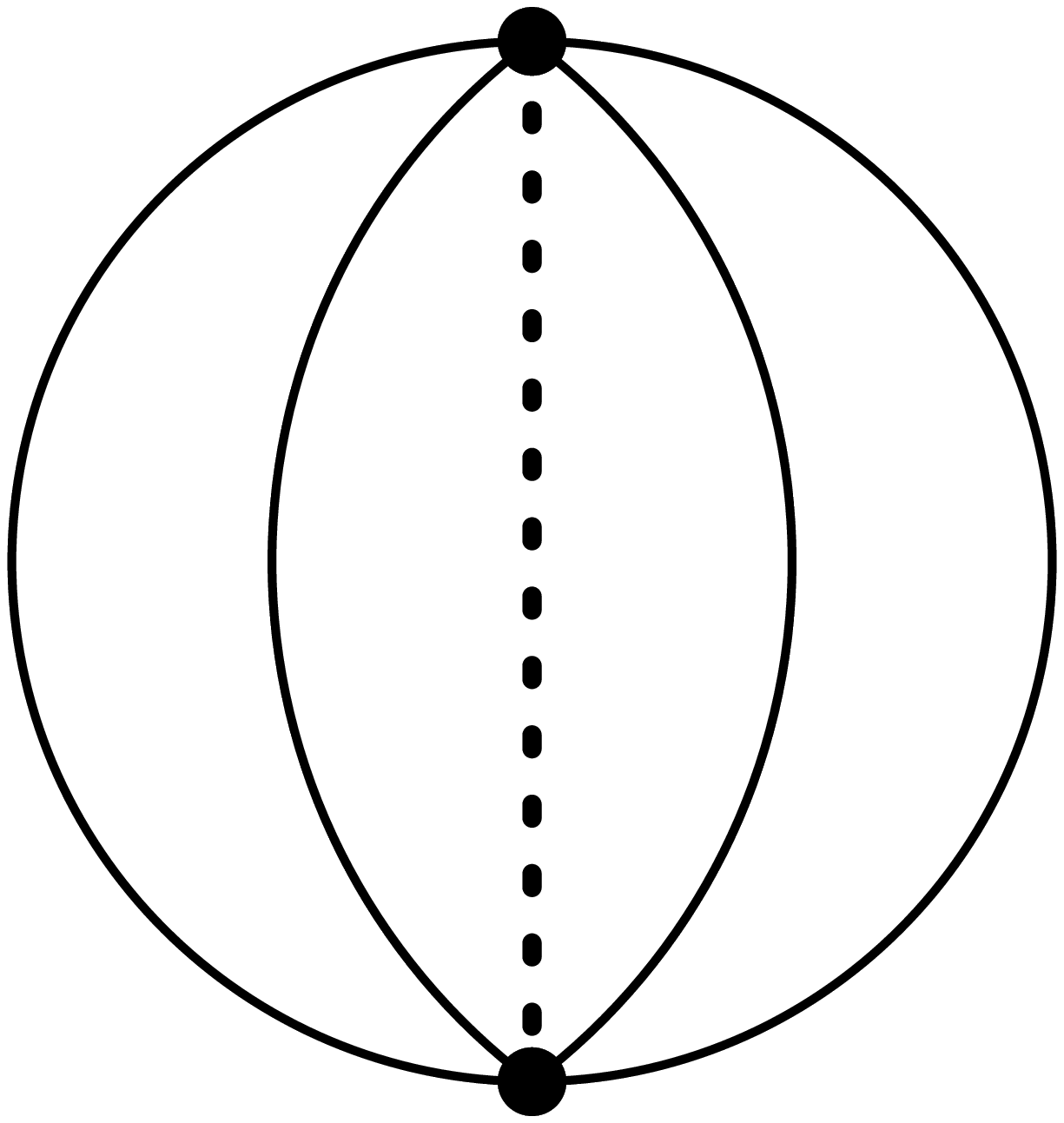}
    $\M{}{5}{4}$
  \end{center}
\end{minipage}
%
\begin{minipage}[b]{2cm}
  \begin{center}
    \includegraphics[height=1.5cm,bb=126 320 460 678]{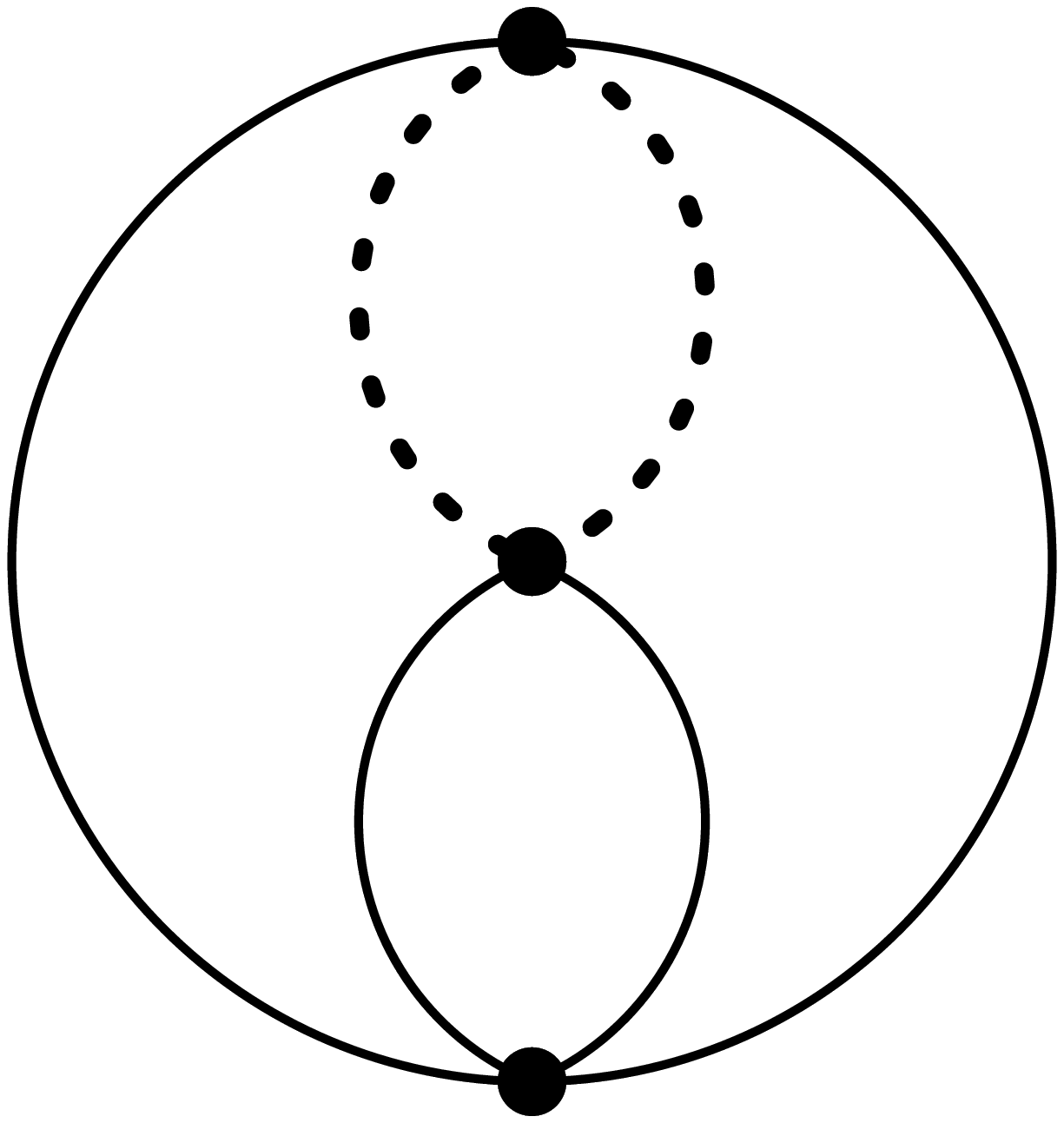}
    $\M{}{6}{2}$
  \end{center}
\end{minipage}
%
%
\begin{minipage}[b]{2cm}
  \begin{center}
    \includegraphics[height=1.5cm,bb=126 320 460 678]{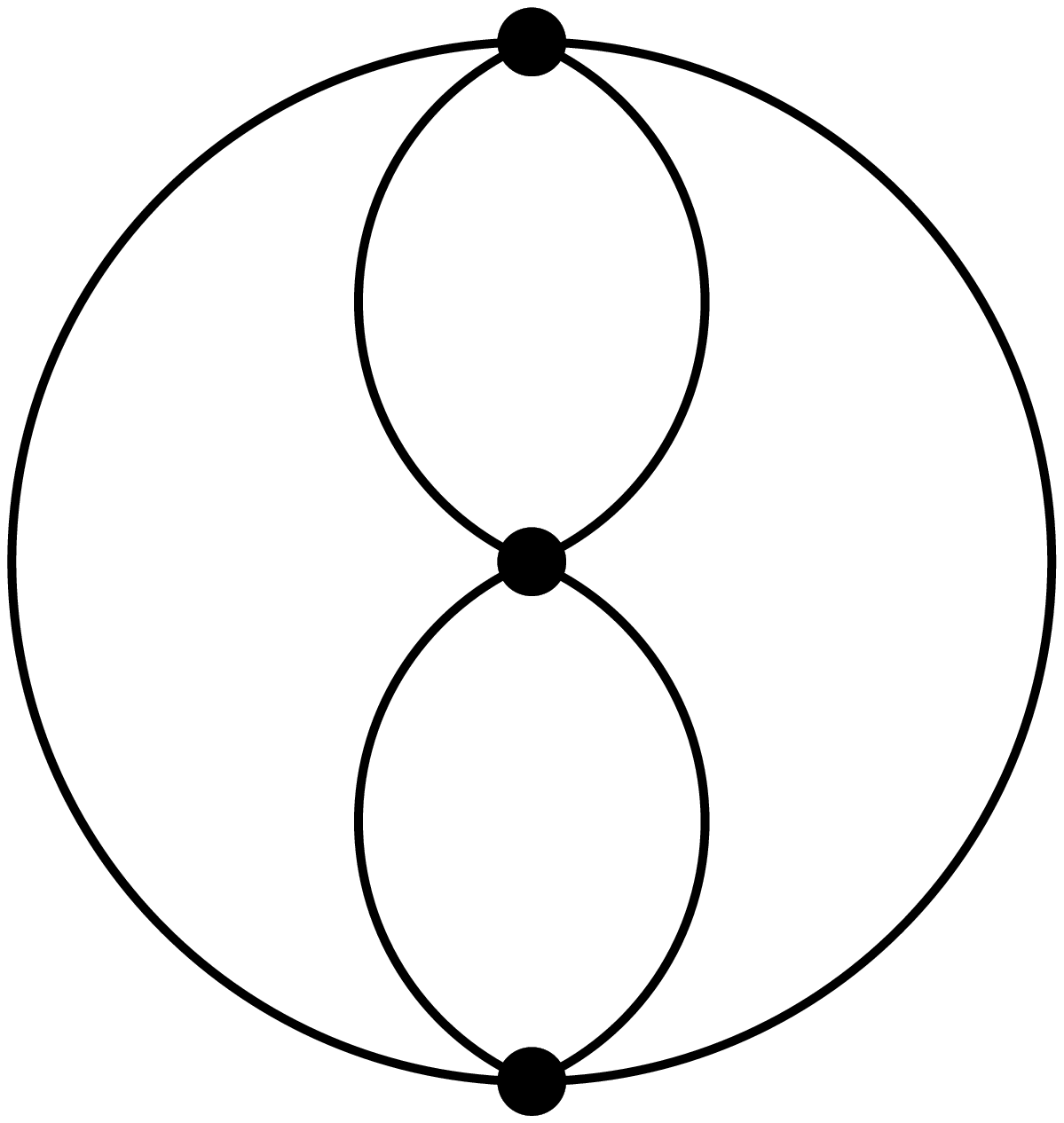}
    $\M{}{6}{1}$
  \end{center}
\end{minipage}
%
%
\begin{minipage}[b]{2cm}
  \begin{center}
    \includegraphics[height=1.5cm,bb=126 332 460 678]{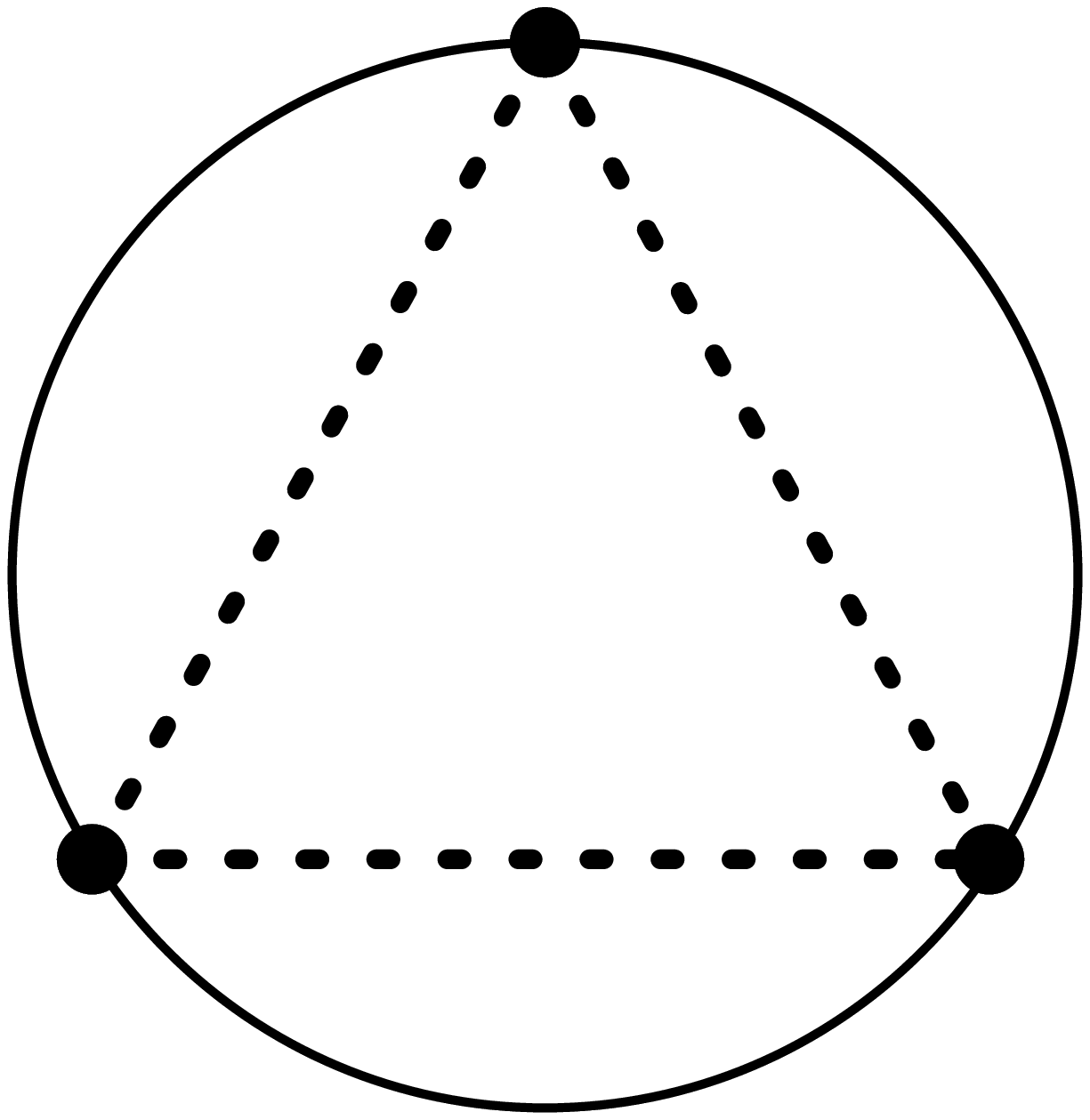}
    $\M{}{6}{4}$
  \end{center}
\end{minipage}
\\
%
%
\begin{minipage}[b]{2cm}
  \begin{center}
    \includegraphics[height=1.5cm,bb=126 332 460 678]{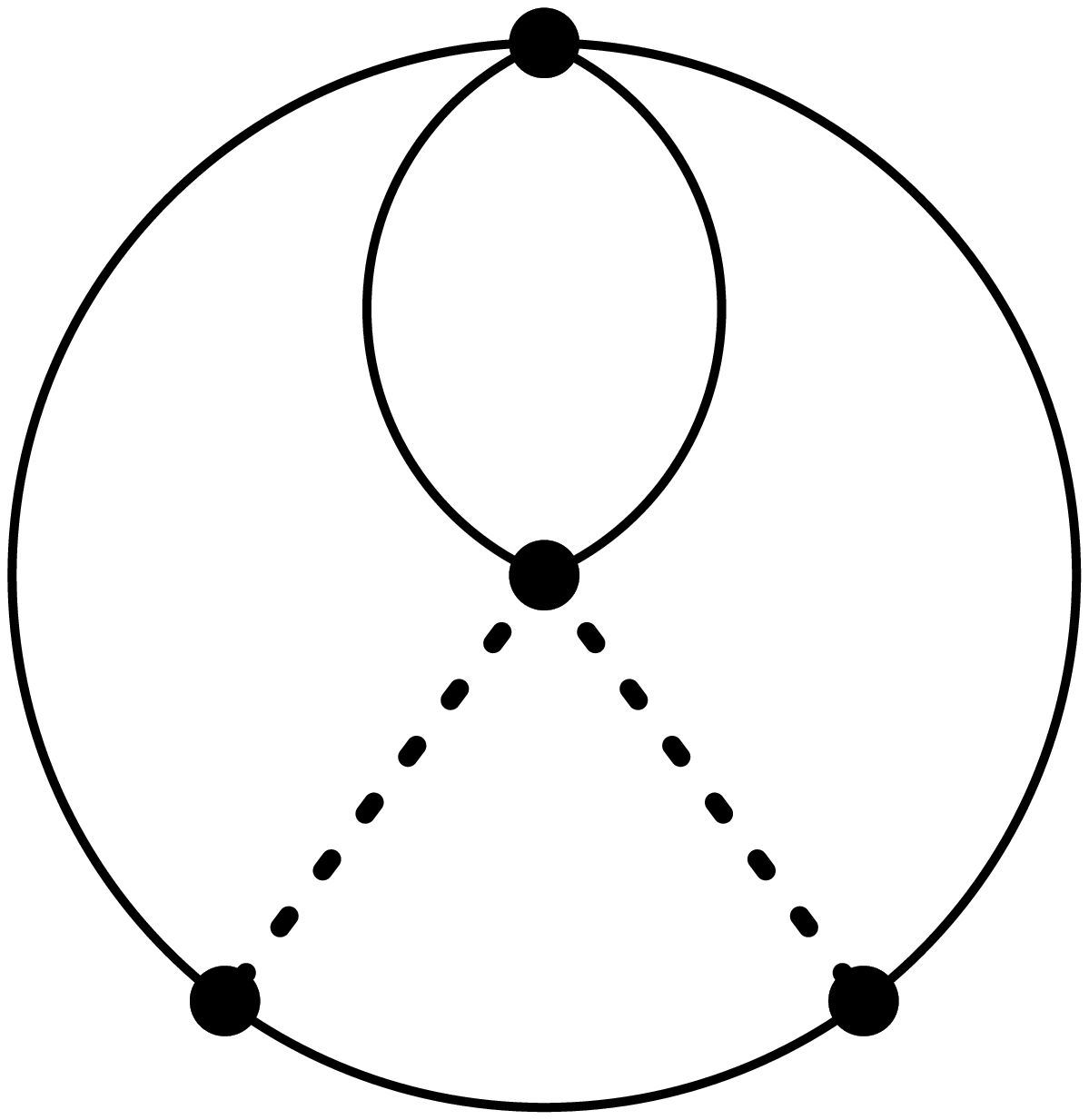}
    $\M{}{7}{1}$
  \end{center}
\end{minipage}
%
%
\begin{minipage}[b]{2cm}
  \begin{center}
    \includegraphics[height=1.5cm,bb=126 332 460 666]{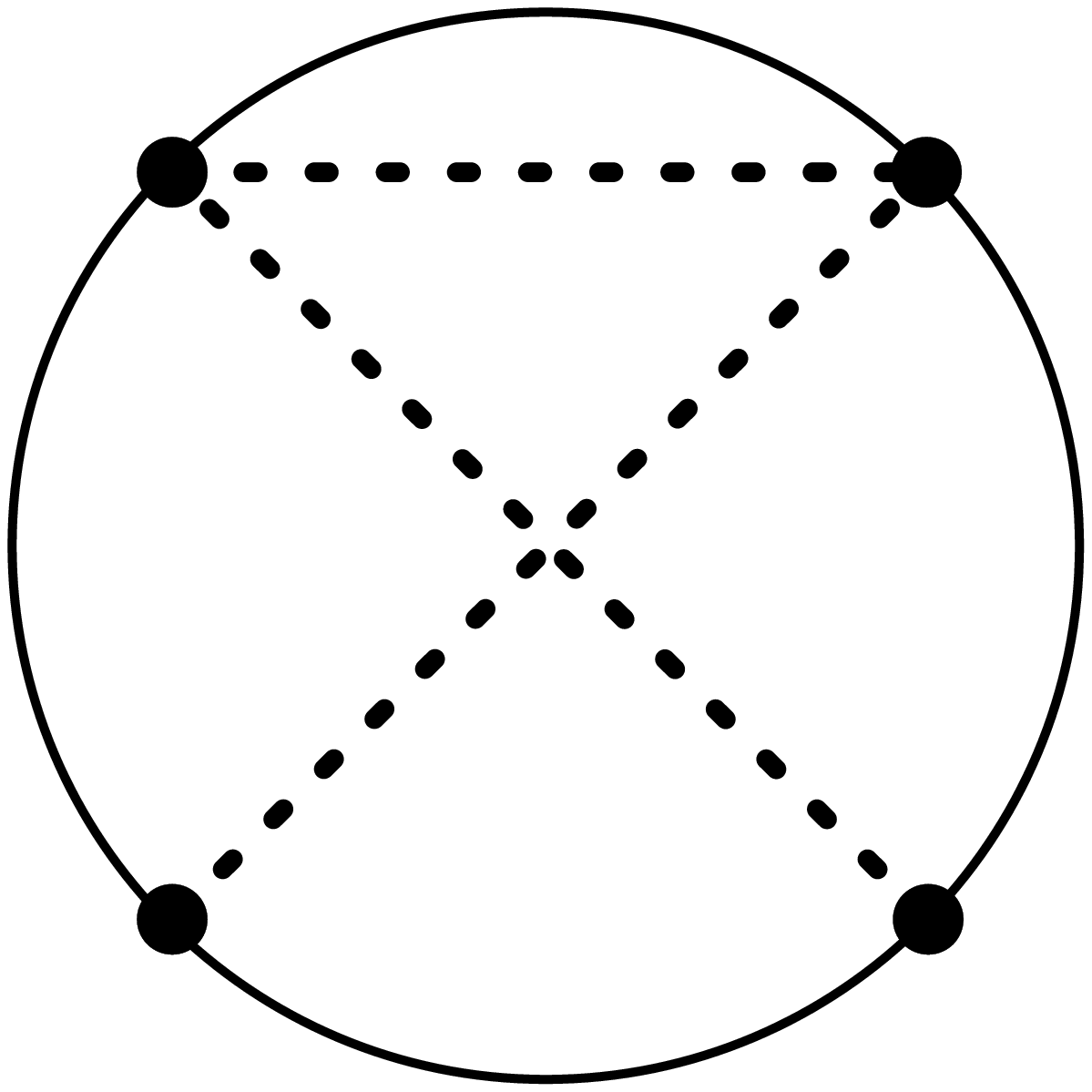}
    $\M{}{7}{2}$
  \end{center}
\end{minipage}
%
%
\begin{minipage}[b]{2cm}
  \begin{center}
    \includegraphics[height=1.5cm,bb=126 332 460 666]{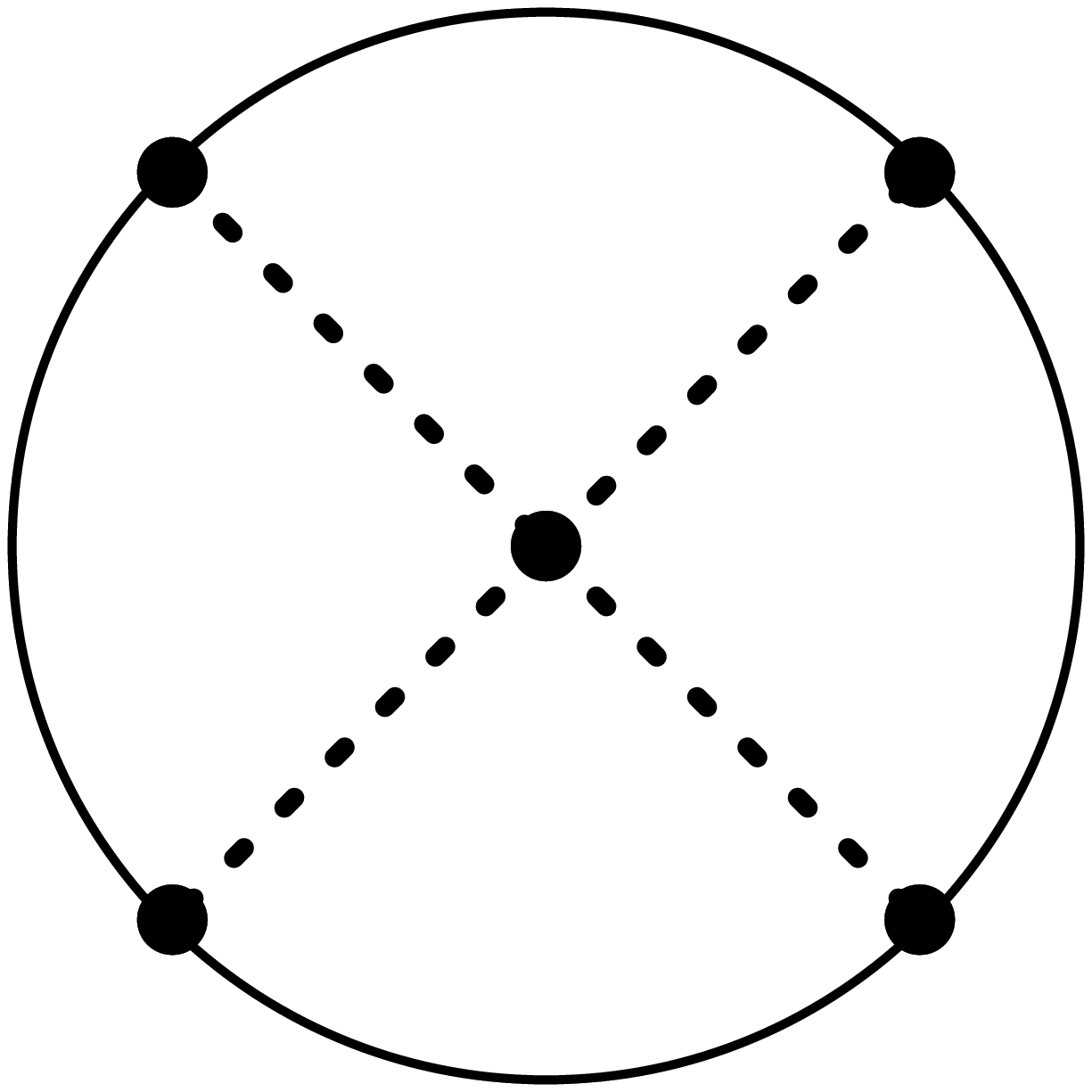}
    $\M{}{8}{1}$
  \end{center}
\end{minipage}
%
%
\begin{minipage}[b]{2cm}
  \begin{center}
    \includegraphics[height=1.5cm,bb=126 320 460 678]{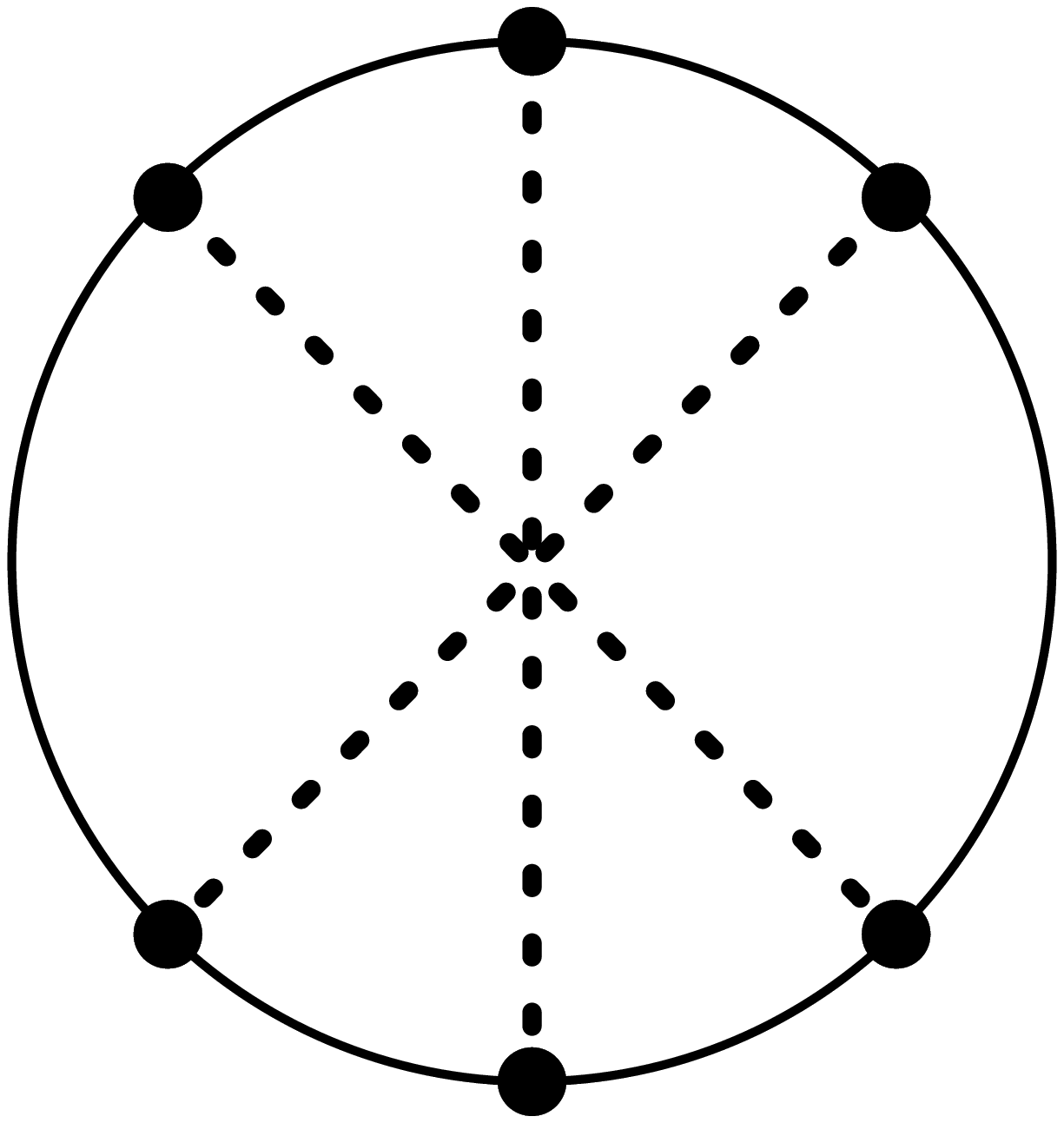}
    $\M{}{9}{1}$
  \end{center}
\end{minipage}
 \end{center}
\vspace{-0.6cm}
\caption{Master integrals in the standard basis. The solid (dashed)
  lines denote massive (massless) propagators.
\label{fig:Stdbasis}}
\end{figure}

In a first step the program {\tt{QGRAF}}~\cite{Nogueira:1991ex} has been
used to generate all necessary diagrams. Subsequently all appearing
integrals have been mapped on a small set of 13 master integrals with
the traditional Integration-By-Parts(IBP) method~\cite{Chetyrkin:1981qh}
in combination with Laporta's
algorithm~\cite{Laporta:1996mq,Laporta:2001dd}. This procedure has been
coded with the help of the programs
{\tt{FORM}}~\cite{Vermaseren:2000nd,Vermaseren:2002rp,Tentyukov:2006ys}
and {\tt{FERMAT}} \cite{Lewis}.  The remaining master integrals are
shown in Fig.~\ref{fig:Stdbasis} in the standard basis. They have first
been determined in Ref.~\cite{Schroder:2005va} with the method of
difference equation~\cite{ Laporta:2000dc,Laporta:2001dd} and
subsequently in Ref.~\cite{Chetyrkin:2006dh}, where the method of
$\varepsilon$-finite basis has been introduced.  Also other
authors~\cite{Laporta:2002pg,Chetyrkin:2004fq,Kniehl:2005yc,Schroder:2005db,Bejdakic:2006vg,Kniehl:2006bf,Kniehl:2006bg}
have contributed in this connection with analytical or numerical
results.

\noindent
The four-loop moment with $n=1$ is known analytically for all four
correlators. For the scalar correlator it is given by
\begin{eqnarray}
\C{1}{0}{3}{0}{s} &=&
   {240320\over 1701}\*\Pa5 
 + {513923\over 5103}\*\Pa4
 + {32995\over 1701}\*\z5
 - {1707578737\over 3265920}\*\z3 
 - {6008\over 5103}\*\Log{2}{5} 
\nonumber\\&&
 + {513923\over 122472}\*\Log{2}{4} 
 + {30040\over 15309}\*\Log{2}{3}\*\pi^2 
 - {513923\over 122472}\*\Log{2}{2}\*\pi^2 
\nonumber\\&&
 + {20122\over 15309}\*\log{(2)}\*\pi^4 
 + {11458913\over 2939328}\*\pi^4
 + {183424051\over 4898880}
,\\
\C{1}{h}{3}{0}{s} &=&
   {13326713\over 34020 }\*\Pa4
 + {4\over 9}\*\z5
 + {609136933177\over 2514758400}\*\z3
 + {13326713\over 816480}\*\Log{2}{4}
\nonumber\\&&
 - {13326713\over 816480}\*\Log{2}{2}\*\pi^2
 - {454390553\over 97977600}\*\pi^4
 + {91614310199\over 3772137600}
,\\
\C{1}{l}{3}{0}{s} &=&
   {14\over 81}\*\Pa4
 + {6278503\over 233280}\*\z3
 + {7\over 972 }\*\Log{2}{4}
 - {7\over 972}\*\Log{2}{2}\*\pi^2
\nonumber\\&& 
 - {35927\over 116640}\*\pi^4
 - {2197597\over 349920 }
,\\
\C{1}{hh}{3}{0}{s} &=&
 - {27479\over 34020}\*\z3
 + {1729337\over 1377810}
,\\
\C{1}{hl}{3}{0}{s} &=&
 - {205\over 324 }\*\Pa4
 - {38171\over 62208}\*\z3
 - {205\over 7776}\*\Log{2}{4}
 + {205\over 7776}\*\Log{2}{2}\*\pi^2 
\nonumber\\&& 
 + {2009\over 186624}\*\pi^4
 + {1937539\over 4199040}
,\\
\C{1}{ll}{3}{0}{s} &=&
   {7867\over 32805}
,\\
\CL{1}{3}{1}{s} &=&
   {197329\over 1296}\*\z3
 - {2573\over 20}
 + \nh\*\left(
   {2541989\over 233280} 
 - {429089\over 31104}\*\z3
       \right) 
\nonumber\\&&
 + \nl\*\bigg(
   {5843\over 1215} 
 - {17939\over 1944}\*\z3
        \bigg) 
 - \nh^2\*\left(
   {60559\over 349920} 
 - {1435\over 5184}\*\z3
         \right)
\nonumber\\&&
 + \nl\*\nh\*\left(
   {1597\over 69984} 
 + {1435\over 5184}\*\z3
             \right) 
 +  \nl^2\*{238\over 1215}
,\\
\CL{1}{3}{2}{s} &=&
   {7381\over 1620}
 - \nh\*{671\over 1215}
 - \nl\*{671\over 1215}
 + \nh^2\*{61\over 3645}
 + \nl\*\nh\*{122\over 3645}
 + \nl^2\*{61\over 3645}
,
\end{eqnarray}
where $\zeta_n$ denotes the Riemann zeta-function and $a_n=\mbox{Li}_n(1/2)$.
The result for the axial-vector correlator is given by
\begin{eqnarray}
\C{1}{0}{3}{0}{a}&=&
 - {3996704\over 25515}\*\Pa5 
 + {5966100779\over 12247200}\*\Pa4 
 + {499588\over 382725}\*\Log{2}{5} 
\nonumber\\&&
 + {5966100779\over 293932800}\*\Log{2}{4}
 - {499588\over 229635}\*\Log{2}{3}\*\pi^2 
 - {5966100779\over 293932800}\*\Log{2}{2}\*\pi^2
\nonumber\\&&
 - {7151983\over 2296350}\*\log{(2)}\*\pi^4
 + {25214645053\over 35271936000}\*\pi^4 
 + {26239187\over 34020}\*\z5
\nonumber\\&&
 - {1910114229901\over 1306368000}\*\z3 
 + {6038304844519\over 5878656000} 
,\\
\C{1}{h}{3}{0}{a}&=&
   {453463328653\over 4115059200} 
 + {920203769\over 680400}\*\Pa4 
 + {920203769\over 16329600}\*\Log{2}{4} 
 - {182\over 27}\*\z5
\nonumber\\&&
 - {920203769\over 16329600}\*\Log{2}{2}\*\pi^2 
 - {6237707419\over 391910400}\*\pi^4 
 + {2277007338343\over 2743372800}\*\z3 
,\\
\C{1}{l}{3}{0}{a}&=&
 -  {33628673\over 839808} 
 - {12907\over 29160}\*\Pa4 
 - {12907\over 699840}\*\Log{2}{4} 
 + {12907\over 699840}\*\Log{2}{2}\*\pi^2 
\nonumber\\&&
 - {16384897\over 16796160}\*\pi^4 
 + {313282529\over 2799360}\*\z3
,\\
\C{1}{hh}{3}{0}{a}&=&
   {15134719\over 16533720}
 - {303799\over 408240}\*\z3
,\\
\C{1}{hl}{3}{0}{a}&=&
   {21592349\over 50388480}
 - {1499\over 3888}\*\Pa4 
 - {1499\over 93312}\*\Log{2}{4} 
 + {1499\over 93312}\*\Log{2}{2}\*\pi^2 
\nonumber\\&&
 + {73451\over 11197440}\*\pi^4 
 - {2539913\over 3732480}\*\z3
,\\
\C{1}{ll}{3}{0}{a}&=&
   {42133\over 196830} 
 - {56\over 405}\*\z3
,\\
\CL{1}{3}{1}{a}&=&
   {96167813\over 311040}\*\z3 
 - {164649889\over 466560 }
 - \nh^2\*\left(
   {806681\over 4199040} 
 - {10493\over 62208}\*\z3
         \right) 
\nonumber\\&&
 - \nl\*\nh\*\left(
   {349337\over 4199040} 
 - {10493\over 62208}\*\z3
             \right) 
 +  \nl^2\*{397\over 3645 }
 + \nh\*\bigg(
   {33280729\over 933120} 
\nonumber\\&&
 - {59917211\over 1866240}\*\z3
       \bigg)
 + \nl\*\left(
   {22748503\over 699840} 
 - {14152979\over 466560}\*\z3
        \right) 
,\\
\CL{1}{3}{2}{a}&=&
 - {1819\over 1620 }
 +  \nh\*{1442\over 3645 }
 + \nl\*{1442\over 3645 }
 - \nh^2\*{13\over 10935 }
\nonumber\\&&
 - \nl\*\nh\*{26\over 10935}
 - \nl^2\*{13\over 10935 }
,\\
\CL{1}{3}{3}{a}&=&
   {7\over 15} 
 -  \nh\*{4\over 27}
 - \nl\*{4\over 27} 
 + \nh^2\*{4\over 405}
 + \nl\*\nh\*{8\over 405}
 + \nl^2\*{4\over 405}
.
\end{eqnarray}
Numerical results for the pseudo-scalar case have first been presented
in Ref.~\cite{Lepage:2008}. The analytical result is given in
Appendix~\ref{app:PseudoVector}, together with the corresponding one for
the vector current correlator, taken from
Ref.~\cite{Chetyrkin:2006xg,Boughezal:2006px}. The coefficients
$\overline{C}_{1}^{\delta}$ $(\delta=s,a,p,v)$ and
$\overline{C}_{2}^{p}$ are listed in numerical form in Table~\ref{tab:1}.
\begin{table}[ht]
\begin{center}
{\footnotesize
\addtolength{\columnsep}{-5ex}
\begin{tabular}{@{\s}c@{\s}|@{\s}c@{\s}|@{\s}r@{\s}r@{\s}r@{\s}r@{\s}r@{\s}r@{\s}r@{\s}r@{\s}r@{\s}r@{\s}}
\hline
\hline
$\delta$ &  $n$ &
  $\overline{C}_n^{(0),\delta}$  & $\overline{C}_n^{(10),\delta}$ & $\overline{C}_n^{(11),\delta}$ &
  $\overline{C}_n^{(20),\delta}$ & $\overline{C}_n^{(21),\delta}$ & $\overline{C}_n^{(22,\delta)}$ &
  $\overline{C}_n^{(30),\delta}$ & $\overline{C}_n^{(31),\delta}$ & $\overline{C}_n^{(32),\delta}$ &
  $\overline{C}_n^{(33),\delta}$
  \\
  \hline
$s$&1&$ 0.8000$&$ 0.6025$&$ 0.0000$&$-7.7402$&$-1.2551$&$ 0.0000$&$-5.4135$&$ 32.8284$&$ 2.6149$&$ 0.0000$\\
\hline
$a$&1&$ 0.5333$&$ 0.8461$&$ 1.0667$&$-1.1071$&$ 1.6925$&$-0.0444$&$-2.4297$&$  5.2981$&$ 0.4406$&$ 0.0321$\\
\hline
$p$&1&$ 1.3333$&$ 3.1111$&$ 0.0000$&$ 0.1154$&$-6.4815$&$ 0.0000$&$-1.2224$&$  2.5008$&$13.5031$&$ 0.0000$\\
$p$&2&$ 0.5333$&$ 2.0642$&$ 1.0667$&$ 7.2362$&$ 1.5909$&$-0.0444$&$ 7.0659$&$ -7.5852$&$ 0.5505$&$ 0.0321$\\
\hline
$v$&1&$ 1.0667$&$ 2.5547$&$ 2.1333$&$ 2.4967$&$ 3.3130$&$-0.0889$&$-5.6404$&$  4.0669$&$ 0.9590$&$ 0.0642$\\
\hline
\hline
\end{tabular}
}
\end{center}
\caption{\label{tab:1} Numerical values for $\overline{C}_{1}^{\delta}$
  $(\delta=s,a,p,v)$ and $\overline{C}_{2}^{p}$ for $n_l=3$ which
  corresponds to the case of the charmed quark.}
\end{table}\\
%
In Appendix~\ref{app:Mom0m1} we provide in addition the expansion
coefficients for $n=0,-1$ . Except for $\overline{C}^{a}_{0}$ they are
only known numerically, however, with high precision.  A completely
analytical result would require the analytical determination of the
constants $\T{54}{2}$, $\T{64}{2}$, $\T{61}{2}$, $\T{62}{3}$,
$\T{72}{1}$, $\T{71}{1}$, $\T{81}{1}$, $\T{91}{1}$, where $\T{n}{i}$ are
the coefficients of the $\varepsilon$-expansion ($\varepsilon=(4-d)/2$)
of the master integrals $T_n$ as defined in Fig.~\ref{fig:Stdbasis} 
\begin{equation}
T_{n}=\sum_{i=n_{min}}^{\infty}\varepsilon^i \T{n}{i}\,.
\end{equation}
\section{Summary and Conclusion\label{sec:DiscussConclude}}
We have computed the lowest coefficients in the low-energy
expansion of the scalar and axial-vector current correlators at
four-loop order in perturbative QCD. All appearing loop-integrals have
been reduced to known master integrals, using Laporta's
algorithm. We also gave the details of the calculation as well as the
analytical results for the moments of the pseudo-scalar
correlator. These results allow to reduce the theoretical error
originating from higher order corrections in the determination of
fundamental constants of QCD, like the strong coupling constant and the
charm-quark mass in the context of lattice calculations.\\

\parindent0cm
{\bf Acknowledgments:}\\
C.S. would like to thank K.G.~Chetyrkin, J.H.~K{\"u}hn and
M.~Steinhauser for their support and many discussions. This research was
supported by U.S. Department of Energy contract No.DE-AC02-98CH10886.
Our results are also available in computer readable form under the URL
{\tt{http://arxiv.org}} by downloading the source of this article.
\begin{appendix}
\section{Moments for the pseudo-scalar and vector\\ correlator\label{app:PseudoVector}}
The analytical result for the first two moments of the pseudo-scalar
correlator presented in numerical from in Ref.~\cite{Lepage:2008} is
given by 
\begin{eqnarray}
\C{1}{0}{3}{0}{p} &=&
   {10304\over 243}\*\Pa5 
 - {130535\over 1458}\*\Pa4 
 - {35189\over 243}\*\z5 
 + {4653637\over 15552}\*\z3 
 - {1288\over 3645}\*\Log{2}{5} 
\nonumber\\&&
 - {130535\over 34992}\*\Log{2}{4} 
 + {1288\over 2187}\*\Log{2}{3}\*\pi^2 
 + {130535\over 34992}\*\Log{2}{2}\*\pi^2
\nonumber\\&& 
 + {10094\over 10935}\*\log{(2)}\*\pi^4 
 - {5686729\over 4199040}\*\pi^4 
 - {3732431\over 34992} 
,\\
\C{1}{h}{3}{0}{p} &=&
 -  {294727\over 2430}\*\Pa4
 + {80\over 9}\*\z5
 - {7656133\over 85050}\*\z3
 - {294727\over 58320}\*\Log{2}{4}
\nonumber\\&&
 + {294727\over 58320}\*\Log{2}{2}\*\pi^2
 + {9593011\over 6998400}\*\pi^4
 - {373843\over 453600 }
,\\
\C{1}{l}{3}{0}{p} &=&
   {115\over 243}\*\Pa4 
 - {157783\over 7776}\*\z3
 + {115\over 5832}\*\Log{2}{4}
 - {115\over 5832}\*\Log{2}{2}\*\pi^2
\nonumber\\&&
 + {115709\over 699840}\*\pi^4
 - {7381\over 5832}
,\\
\C{1}{hh}{3}{0}{p} &=&
   {403\over 630}\*\z3
 - {9493\over 25515}
,\\
\C{1}{hl}{3}{0}{p} &=&
 -  {4\over 9}\*\Pa4 
 - {1\over 54}\*\Log{2}{4} 
 + {1\over 54}\*\Log{2}{2}\*\pi^2 
 + {49\over 6480}\*\pi^4 
 + {35\over 96}\*\z3
 - {3457\over 11664}
,\\
\C{1}{ll}{3}{0}{p} &=&
   {277\over 729}
,\\
\CL{1}{3}{1}{p} &=&
 - {71203\over 864}\*\z3
 + {143465\over 1296}
 - \nh\*\left(
   {12439\over 1944} 
 - {2315\over 1296}\*\z3
        \right)
 - \nl\*\bigg(
   {17191\over 1944} 
\nonumber\\&&
 - {6473\over 1296}\*\z3
        \bigg)
 + \nh^2\*\left(
   {14\over 243} 
 + {7\over 36}\*\z3
         \right) 
 + \nl\*\nh\*\left(
   {64\over 243} 
 + {7\over 36}\*\z3
             \right) 
 +  \nl^2\*{50\over 243}
,\\
\CL{1}{3}{2}{p} &=&
   {847\over 36}     
 - \nh\*{77\over 27}
 - \nl\*{77\over 27}
 + \nh^2\*{7\over 81} 
 + \nl\*\nh\*{14\over 81}
 +  \nl^2\*{7\over 81} 
,\\
\C{2}{0}{3}{0}{p} &=&
  {278048\over 945}\*\Pa5 
- {3509250197\over 4082400}\*\Pa4 
- {45178393\over 34020}\*\z5 
+ {2871407869129\over 1306368000}\*\z3 
\nonumber\\&&
- {34756\over 14175}\*\Log{2}{5} 
- {3509250197\over 97977600}\*\Log{2}{4} 
+ {3509250197\over 97977600}\*\Log{2}{2}\*\pi^2 
\nonumber\\&&
+ {34756\over 8505}\*\Log{2}{3}\*\pi^2 
+ {180277\over 28350}\*\log{(2)}\*\pi^4 
- {94380515779\over 11757312000}\*\pi^4 
\nonumber\\&&
- {509351043139\over 653184000 }
,\\
\C{2}{h}{3}{0}{p} &=&
- {978527581\over 680400 }\*\Pa4
+ {62\over 3}\*\z5
- {8948001289387\over 10059033600 }\*\z3
- {978527581\over 16329600 }\*\Log{2}{4}
\nonumber\\&&
+ {978527581\over 16329600 }\*\Log{2}{2}\*\pi^2
+ {6586806599\over 391910400 }\*\pi^4
- {351736938533\over 3017710080 }
,\\
\C{2}{l}{3}{0}{p} &=&
- {493\over 3240 }\*\Pa4
- {151413217\over 933120 }\*\z3
- {493\over 77760 }\*\Log{2}{4}
+ {493\over 77760 }\*\Log{2}{2}\*\pi^2
\nonumber\\&&
+ {1044179\over 622080 }\*\pi^4
+ {8402929\over 279936 }
,\\
\C{2}{hh}{3}{0}{p} &=&
  {36809\over 136080 }\*\z3
- {1699529\over 5511240 }
,\\
\C{2}{hl}{3}{0}{p} &=&
- {179\over 1296 }\*\Pa4
+ {17839\over 1244160 }\*\z3
- {179\over 31104 }\*\Log{2}{4}
\nonumber\\&&
+ {179\over 31104 }\*\Log{2}{2}\*\pi^2
+ {8771\over 3732480 }\*\pi^4
- {1951867\over 16796160 }
,\\
\C{2}{ll}{3}{0}{p} &=&
- {56\over 405 }\*\z3
+ {15511\over 65610 }
,\\
\CL{2}{3}{1}{p} &=&
- {54646039\over 103680 }\*\z3
+ {97431227\over 155520 }
- \nh\*\left(
       {18258607\over 311040 }
      -{30278653\over 622080 }\*\z3
      \right)
\nonumber\\&&
- \nl\*\left(
       {13928429\over 233280 }
      -{7668337\over 155520 }\*\z3
       \right)
- \nh^2\*\left(
         {39137\over 1399680 }
        -{1253\over 20736 }\*\z3
        \right)
\nonumber\\&&
+ \nh\*\nl\*\left(
         {55711\over 1399680 }
        +{1253\over 20736 }\*\z3
            \right)
+ \nl^2\*{247\over 3645 }
,\\
\CL{2}{3}{2}{p} &=&
  {257\over 540 }
- \nh\*{136\over 1215 }
- \nl\*{136\over 1215 }
+ \nh^2\*{119\over 3645} 
+ \nh\*\nl\*{238\over 3645 }
+ \nl^2\*{119\over 3645 }
,\\
\CL{2}{3}{3}{p} &=&
  {7\over 15 }
- \nh\*{4\over 27 }
- \nl\*{4\over 27 }
+ \nh^2\*{4\over 405 }
+ \nh\*\nl\*{8\over 405 }
+ \nl^2\*{4\over 405 }
.
\end{eqnarray}
The first moment of the vector current correlator has been obtained in
Ref.~\cite{Chetyrkin:2006xg,Boughezal:2006px}
\begin{eqnarray}
\C{1}{0}{3}{0}{v} &=&
- {1019840\over 5103 }\*\Pa5
- {84951877\over 306180 }\*\Pa4
- {3655\over 10206 }\*\z5
+ {17554601717\over 32659200 }\*\z3
+ {25496\over 15309 }\*\Log{2}{5}
\nonumber\\&&
- {84951877\over 7348320 }\*\Log{2}{4}
- {127480\over 45927 }\*\Log{2}{3}\*\pi^2
+ {84951877\over 7348320 }\*\Log{2}{2}\*\pi^2
\nonumber\\&&
- {359687\over 229635 }\*\log{(2)}\*\pi^4
- {2653167371\over 881798400 }\*\pi^4
- {5397779543\over 146966400 }
,\\
\C{1}{h}{3}{0}{v} &=&
- {1394804\over 8505 }\*\Pa4
+ {128\over 27 }\*\z5
- {95617883401\over 943034400 }\*\z3
- {348701\over 51030 }\*\Log{2}{4}
\nonumber\\&&
+ {348701\over 51030 }\*\Log{2}{2}\*\pi^2
+ {1447057\over 765450 }\*\pi^4
- {27670774337\over 1414551600 }
,\\
\C{1}{l}{3}{0}{v} &=&
- {4793\over 7290 }\*\Pa4
- {48350497\over 1399680 }\*\z3
- {4793\over 174960 }\*\Log{2}{4}
\nonumber\\&&
+ {4793\over 174960 }\*\Log{2}{2}\*\pi^2
+ {372689\over 839808 }\*\pi^4
- {9338899\over 2099520 }
,\\
\C{1}{hh}{3}{0}{v} &=&
- {3287\over 7290 }\*\z3
+ {163868\over 295245 }
,\\
\C{1}{hl}{3}{0}{v} &=&
- {116\over 243 }\*\Pa4
- {38909\over 58320 }\*\z3
- {29\over 1458 }\*\Log{2}{4}
\nonumber\\&&
+ {1421\over 174960 }\*\pi^4
+ {29\over 1458 }\*\Log{2}{2}\*\pi^2
+ {262877\over 787320 }
,\\
\C{1}{ll}{3}{0}{v} &=&
  {42173\over 98415 }
- {112\over 405 }\*\z3
,\\
\CL{1}{3}{1}{v} &=&
  {7236859\over 38880 }
- {10589033\over 77760 }\*\z3
- \nh\*\left(
       {520823\over 34992 }
      -{1049579\over 116640 }\*\z3
       \right)
\nonumber\\&&
- \nl\*\left(
       {1103117\over 58320 }
      -{1305359\over 116640 }\*\z3
      \right)
- \nh^2\*\left(
       {14483\over 65610 }
      -{203\over 972 }\*\z3
        \right)
\nonumber\\&&
- \nh\*\nl\*\left(
            {3779\over 65610 }
           -{203\over 972 }\*\z3
           \right)
+ \nl^2\*{1784\over 10935 }
,\\
\CL{1}{3}{2}{v} &=&
- {451\over 405 }
+ \nh\*{1574\over 3645 }
+ \nl\*{1574\over 3645 }
+ \nh^2\*{236\over 10935 }
+ \nh\*\nl\*{472\over 10935 }
+ \nl^2\*{236\over 10935 }
,\\
\CL{1}{3}{3}{v} &=&
  {14\over 15 }
- \nh\*{8\over 27 } 
- \nl\*{8\over 27 }
+ \nh^2\*{8\over 405 }
+ \nh\*\nl\*{16\over 405 }
+ \nl^2\*{8\over 405 }
.
\end{eqnarray}
%
%
%
%
%
%
%
%
%
%
\section{Moments for {$\mathbf{n=-1,0}$}\label{app:Mom0m1}}
The moments for $n=-1$ and $n=0$ exhibit an overall UV divergence. The
corresponding $1/\varepsilon$-poles are dropped by definition in the 
$\overline{\mbox{MS}}$-scheme. The finite parts are given by
\begin{eqnarray}
\overline{C}_{-1}^{(3),s}&=&
- 325.6276432 
+ 16.39537650\*\nh 
+ 19.76434509\*\nl 
\nonumber\\&&
- 1.670198265\*\nh^2 
- 0.9856898698\*\nl\*\nh
+ 0.7103788267\*\nl^2 
,\\
\overline{C}_{0\phantom{+}}^{(3),s}&=&
- 82.24477459 
+ 14.55555905\*\nh 
+ 22.86448444\*\nl 
\nonumber\\&&
- 0.6046449347\*\nh^2 
- 1.620853371\*\nl\*\nh
- 1.047986065\*\nl^2 
,\\
\overline{C}_{-1}^{(3),p}&=&
- 358.5973626 
+ 30.02655475\*\nh 
+ 35.65136262\*\nl 
\nonumber\\&&
- 1.485773643\*\nh^2 
- 2.003379176\*\nl\*\nh
- 0.2645810947\*\nl^2 
,\\
\overline{C}_{0\phantom{+}}^{(3),p}&=&
- 25.62696915 
+ 8.147150256\*\nh 
+ 6.938402913\*\nl 
\nonumber\\&&
+ 0.08246295965\*\nh^2 
- 0.5283433562\*\nl\*\nh
- 0.4972293123\*\nl^2 
,\\
\overline{C}_{-1}^{(3),a}&=&
  25.62696915 
- 8.147150256\*\nh 
- 6.938402913\*\nl 
\nonumber\\&&
- 0.08246295965\*\nh^2 
+ 0.5283433562\*\nl\*\nh
+ 0.4972293123\*\nl^2 
,\\
\C{0}{0}{3}{0}{a}&=&
 - {59584\over 729}\*\Pa5 
 + {362533\over 4374}\*\Pa4 
 + {668057\over 5832}\*\z5
 - {29132419\over 233280}\*\z3 
\nonumber\\&&
 + {7448\over 10935}\*\Log{2}{5} 
 + {362533\over 104976}\*\Log{2}{4} 
 - {7448\over 6561}\*\Log{2}{3}\*\pi^2 
\nonumber\\&&
 - {362533\over 104976}\*\Log{2}{2}\*\pi^2 
 - {35584\over 32805}\*\log{(2)}\*\pi^4 
\nonumber\\&&
 - {309377\over 2519424}\*\pi^4 
 + {402928129\over 4199040} 
,\\
\C{0}{ll}{3}{0}{a}&=&
 -  {31\over 162}\*\z3
 - {1655\over 69984 }
,\\
\C{0}{hl}{3}{0}{a}&=&
   {2\over 27}\*\Pa4 
 - {1919\over 2592}\*\z3
 + {1\over 324}\*\Log{2}{4} 
 - {1\over 324}\*\Log{2}{2}\*\pi^2 
\nonumber\\&&
 - {49\over 38880}\*\pi^4 
 + {4255\over 8748} 
,\\
\C{0}{hh}{3}{0}{a}&=&
 -  {913\over 1134}\*\z3
 + {317311\over 489888 }
,\\
\C{0}{l}{3}{0}{a}&=&
 - {1025\over 729}\*\Pa4 
 + {310001\over 23328}\*\z3 
 - {1025\over 17496}\*\Log{2}{4} 
 + {1025\over 17496}\*\Log{2}{2}\*\pi^2 
\nonumber\\&&
 - {18635\over 419904}\*\pi^4 
 - {339551\over 139968} 
,\\
\C{0}{h}{3}{0}{a}&=&
   {859259\over 7290}\*\Pa4 
 - {85\over 27}\*\z5
 + {12520907\over 145800}\*\z3 
 + {859259\over 174960}\*\Log{2}{4} 
\nonumber\\
&-&{859259\over 174960}\*\Log{2}{2}\*\pi^2 
 - {28913567\over 20995200}\*\pi^4 
 - {556867\over 1166400} 
,
\end{eqnarray}
where for compactness $\mu=m$ has been chosen.  The logarithms
$\log(\mu^2/m^2)$ can be reconstructed with the help of the RGE.
%

\end{appendix}

\end{document}